# From domain walls and the stripe phase to full suppression of charge density wave in the superconducting 1T-Ti$_{1-x}$Ta$_x$Se$_2$

Q. Hu[1,*], R. Venturini[1,2,*,†], Y. Vaskivskyi[1,2], J. Lipič[2], Z. Jagličić[3,4], and D. Mihailovic[1,2,5]

[1]Department of Complex Matter, Jozef Stefan Institute, Jamova 39, 1000 Ljubljana, Slovenia

[2]Faculty for Mathematics and Physics, University of Ljubljana, Jadranska 19, 1000 Ljubljana, Slovenia

[3]Faculty of Civil and Geodetic Engineering, University of Ljubljana, Jamova cesta 2, 1000 Ljubljana, Slovenia

[4]Institute of Mathematics, Physics and Mechanics, Jadranska 19, 1000 Ljubljana, Slovenia

[5]CENN Nanocenter, Jamova 39, 1000 Ljubljana, Slovenia

## Abstract

1T-TiSe$_2$ hosts a 2×2×2 charge density wave (CDW) that is known to form the state with localized domains separated by the domain walls upon Cu intercalation. The CDW state with the domain wall network has attracted significant interest due to its coexistence with superconductivity. Here we present a scanning tunneling microscopy, transport and magnetic susceptibility study of 1T-Ti$_{1-x}$Ta$_x$Se$_2$. Ta substitution for Ti atoms allows us to perform experiments over the wide range of doping ($0 \leq x \leq 0.2$), providing access to a significantly broader phase diagram than Cu intercalation experiments. At x = 0.02, we observe a complex network of domains and domain walls. We identify two distinct types of domain walls and show their structure with atomic resolution. Additionally, an elusive symmetry-breaking stripe CDW is found at the light substitution of x = 0.02. We also measure highly substituted x = 0.2 crystals that are superconducting despite the full collapse of the CDW order. Our results uncover rich CDW physics in Ta-substituted 1T-TiSe$_2$ crystals and illuminate the interplay between the CDW and superconductivity.

## Introduction

Charge density wave (CDW) phases observed in two-dimensional materials are still not fully understood despite the work on these compounds since the 1970's which particularly intensified in the last decade. Control over such phases has been demonstrated by chemical doping [1], pressure [2,3], gating [4–6], ultrafast optical [7], and electrical pulse experiments [8,9]. Furthermore, when the CDW order is partially suppressed, the superconducting state commonly emerges [2,10–12], indicating an interplay between CDW and superconductivity that is yet to be fully explored and understood.

The pristine 1T-TiSe$_2$ is a two-dimensional transition metal dichalcogenide [Fig. 1(a)] that undergoes a second-order phase transition from a metallic to a commensurate 2×2×2 CDW at around 200 K [13]. Despite significant efforts in recent years the origin of the CDW has not yet been ambiguously determined with both electron-phonon coupling and excitonic correlations considered to be important [14–17].

The discovery of superconductivity in Cu intercalated 1T-TiSe$_2$ crystals reignited interest in this compound [12]. By increasing the level of Cu intercalation, the long-range commensurate CDW order breaks up into domains separated by domain walls [18–23]. STM studies have revealed that the commensurate domains are separated by π phase-shifted domain walls [18,20]. The phase diagram of the Cu intercalated samples has been expanded further with the observation of the stripe CDW at low intercalation values, where the domain walls are absent [21]. The synthesis of the Cu intercalated crystal allows for an intercalation level of only up to around x ~ 0.11 at which the CDW domain state is still present inside the material [12,19]. As the onset of superconductivity appears to coincide with the formation of the domain walls [19], CDW discommensurations are likely crucial for the promotion of superconductivity [19,23,24]. Some works have argued that there is even a cooperative coexistence between the CDW discommensurations and the superconductivity [23,24].

---

[*] These authors contributed equally to this work

[†] rok.venturini@ijs.si

In this work, we focus on Ta-substituted 1T-TiSe$_2$ samples which, unlike Cu$_x$TiSe$_2$ crystals, have no limitation on doping level which allows us to study the full phase diagram of both domain wall state in lightly substituted samples as well as fully suppressed CDW at a high level of substitution. Ta-substituted samples are interesting as Ta atoms have 5 valence electrons which means that Ta is expected to donate electrons to the conduction band of 1T-TiSe$_2$, similar to Cu intercalated samples. Furthermore, our goal is to study in detail the domain wall structure which is possibly relevant for the emergence of superconductivity. Additionally, the elusive symmetry-breaking stripe phase has so far been reported only in Cu$_{0.01}$TiSe$_2$ samples which do not exhibit domain walls. This begs the question, whether the stripe phase could be stabilized in 1T-TiSe$_2$ with the other other tunning parameters as well.

Our STM measurements of lightly substituted x = 0.02 samples reveal domains and domain walls similar to Cu intercalated 1T-TiSe$_2$ crystals [18,20,21,25]. We emphasize that 2×2×2 CDW geometry in 1T-TiSe$_2$ allows for the formation of two distinct types of domain walls and we show their coexistence on both large-scale scans as well as their detailed atomically resolved image. In x = 0.02 crystal we additionally observe the elusive stripe phase, which has to our knowledge been reported only once [21]. The highly-substituted x = 0.2 samples have fully suppressed CDW order with the superconducting state that has higher critical temperature than x = 0.02 samples.

## Results

High-quality single crystal samples of 1T-Ti$_{1-x}$Ta$_x$Se$_2$ were grown by the chemical vapor transport (CVT) technique with iodine as a transport agent. To avoid significant Ti self-doping, the crystals were grown at a low temperature of 650 °C [13]. For pristine samples grown at 650 °C it has been shown that the 0.35% density of Ti defect does not affect the long-range CDW order [25]. The STM/STS results presented here were measured using Pt-Ir tips prepared on the Au surface with the sample grounded and the bias voltage applied to the tip. The 1T-Ti$_{1-x}$Ta$_x$Se$_2$ single crystals were cleaved in an ultra-high vacuum at room temperature before being transferred to the sample stage.

Figure 1(b) shows the STM image of pristine 1T-TiSe$_2$ obtained at $T$ = 4 K. A clear 2×2 CDW pattern with well-defined CDW superlattice peaks in the Fourier transform [Fig. 1(b) inset] is observed. We performed resistance measurements of pristine, x = 0.02 and x = 0.2 samples down to 10 K [Fig. 1(c)]. The resistance measurements show that the CDW transition is largely, but not fully suppressed for x = 0.02, while any sign of a phase transition is absent for the x = 0.2 sample [see inset to Fig. 1(c)]. SQUID magnetization measurements shown in Fig. 1(d) indicate that x = 0.2 samples are superconducting with a critical temperature of 2.0 K. This matches the observations in Ref. [26], where the superconducting temperature of x = 0.2 sample was found to be 2.2 K. The critical temperature of the x = 0.02 sample was reported to be 1.6 K [26], which is below our lowest achievable temperature. It has also been shown that Ta substitution influences lattice parameters: the in-plane lattice parameter *a* shrinks, while the out-of-plane lattice parameter *c* expands [inset Fig. 1(d)] [26].

First, we focus on lightly substituted x = 0.02 sample. The STM image acquired with a small positive tip voltage shown in Fig. 2(a) reveals a very complex and inhomogeneous electronic structure. We see a collapse of long-range CDW order as the state is broken into commensurate CDW domains of different shapes and sizes. The domains are separated either by regions of strongly suppressed CDW order or by the domain walls. Superficially the state looks similar to the $Cu_xTiSe_2$ [18,20] and highly self-doped 1T-$TiSe_2$ crystals [27].

Additionally, we observe atomic-sized triangular features. These are similar to previously observed Ti defects in self-doped crystals. The Ti self-doping defects are significantly more diluted with a density of 0.35% for crystals grown at 650°C [25]. In Fig. 2(a) we count 145 of such defects which corresponds to a density of 1.7% which is in good agreement with the 2% substitution of the x = 0.02 sample. Therefore, the STM image in Fig. 2(a) is representative of the bulk sample and the spatial distribution of these defects gives us an indication of the nanoscale substitution inhomogeneity.

Fig. 2(b) shows the same area as in Figure 2(a), however, measured at a negative tip voltage with tunneling to the unoccupied states. Black and blue dashed lines indicate domain walls. Recent STM studies have emphasized that the domain walls appear as a π-phase shift in CDW order that is perpendicular with respect to the CDW direction [18,20]. We indicate this kind of domain wall with a black dashed line. However, as schematically shown in Fig. 2(d), the in-plane geometry of a 2×2 CDW on a triangular lattice allows for another type of domain wall that we call here π* domain wall (blue dashed line) where a phase shift occurs along the CDW direction. As such domain walls are less frequent and less obvious they are easy to overlook.

As we have only seen the more evident π shifted domain walls being documented in the literature [18,20,21,28,29], we present both types of domain walls in detail with atomically resolved images in Fig. 2(e). Large white circles indicate CDW maxima and are to be used as a guide to the eye for the phase slip between the two domains. Line scans in Fig. 2(f) show height profiles across the π and π* domain walls respectively. By analyzing images of STM images of domain walls in 1T-$TiSe_2$ due to Cu intercalation from Ref. [21], we indeed also find a π* domain wall present that was not documented (see Fig. S1). Given that π* domain walls are observed in both Ta substituted and Cu intercalated $TiSe_2$ crystals, we suspect that both types of domain walls are likely present also seen in other instances of domain wall networks in 1T-$TiSe_2$ crystals.

Fig. 3(a) shows the STM image of x = 0.02 acquired at 77 K showing the appearance of a symmetry-broken stripe CDW (1$Q$) domain. Due to both 3$Q$ and 1$Q$ CDW orders appearing along the same scan lines, we can rule out that the observed stripe phase is a tip artifact. The Fourier transforms [Fig. 3(b) and 3(c)] reveal that the $q_{1Q}$ component of the stripe phase matches one of the three $q_{3Q}$ components of the 3Q phase. We observed the stripe phase occasionally at 77 K, however, we were not able to find it at 4 K. As the stripe phase is observed very rarely, we can't rule out its existence also at low temperatures.

So far the stripe phase has been reported only once, with the discovery being made by an STM study in a Cu-intercalated 1T-$Cu_{0.01}TiSe_2$ [21]. Interestingly, the stripe phase reported in Ref. [21] was observed at low values of Cu intercalation which are too low for the formation of domain walls. This differs from our experiment where we see that the stripe phase can coexist with the domain wall network.

A similar stripe phase has also been observed in 2H-NbSe$_2$ [30]. In this material, the stripe phase has been observed in the area where the sample is strained, giving rise to the interpretation that the formation of the 1Q phase is driven by strain. Ta substitution of 1T-TiSe$_2$ is known to change the lattice spacing of 1T-TiSe$_2$ as shown in the inset of Fig 1(d). As substitution is not homogenous at the atomic scale, it seems possible, that a certain distribution of Ta atoms creates the strain environment suitable for the stripe phase formation. An alternative viewpoint of how conduction band doping could provide conditions for the formation of the stripe phase was recently proposed [31]. Given that the stripe phase is very rarely observed in either doped or intercalated 1T-TiSe$_2$ samples, it appears that the range of conditions that favor the formation of stripe phase formation is very narrow.

Next, we focus on the superconducting x = 0.2 sample which does not show any signature of a phase transition to the CDW state in transport measurement as shown in Figure 1(c). STM experiments with both negative [Fig. 4(a)] and positive bias voltage [Fig. 4(b)] do not show CDW order, which is further confirmed by the FFT image [Fig. 4(c)]. The non-uniform electronic background is likely due to microscopically nonhomogeneous Ta substitution. The STM measurements were performed at both 77 K and 4 K and on many different spots of the bulk crystal with the same results. Our observations are somewhat different from the previous results characterized by the electron diffraction where CDW diffraction spots have been observed at 89 K even for x = 0.2 crystals, while full suppression of CDW was observed only above x = 0.3 [26]. Given that the areas with domain walls and suppressed CDW order populate a significant portion of the sample already at the light substitution of x = 0.02, our observation of full suppression of the CDW at x = 0.2 is not surprising. For our samples both transport and STM data are consistent in not showing any signature of the CDW state being present in the x = 0.2 crystal. As indicated in Fig. 4(d), tunneling spectroscopy shows an increase in the density of states around the Fermi level compared to the x = 0.02 crystal.

After the superconductivity in Cu intercalated 1T-TiSe$_2$ was discovered, it has been proposed that superconductivity is possibly not only related to the incommensurate CDW order [6,19,23], but potentially also enhanced by it in a cooperation scenario [23,24]. This proposal was supported by the observation in high-pressure experiments where a superconducting dome is limited to an incommensurate CDW region. Our experiments show a different phase diagram in which we see further enhancement of superconductivity as the CDW order is fully suppressed: the x = 0.2 sample is superconducting with a higher critical temperature compared to the x = 0.02 sample with CDW order and domain walls. By using a high level of Ta substitution, we see that the superconducting state in 1T-TiSe$_2$ can develop independently of CDW order. Superconductivity is possibly promoted by the enhanced density of states at the Fermi level as the CDW order is suppressed either inside the domain walls or throughout the whole sample.

## Conclusion

In summary, we conducted STM, electrical transport, and magnetic susceptibility measurements on 1T-Ti$_{1-x}$Ta$_x$Se$_2$ crystals. For x = 0.02 samples, we observe a complex network of distinct π and π* domain walls. Furthermore, at 77 K patches of symmetry broken stripe phase are found to coexist with a domain state. This suggests that modest Ta substitution offers a platform for the rich physics of coexisting CDW states. Microscopic understanding of how long-range order is broken could also be instrumental for understanding the metastable state that forms as a response to pulsed excitations [32].

Ta-substitution has the advantage of giving us access to a broad phase diagram of the 1T-Ti$_{1-x}$Ta$_x$Se$_2$. In highly substituted samples we observe a further enhancement of superconductivity compared to the lightly substituted crystals. As the CDW order in the x = 0.2 samples is fully suppressed, our results are compatible with the scenario of competition between the superconductivity and CDW order in this compound.

## Acknowledgments

The authors acknowledge the financial support of the Slovenian Research and Innovation Agency (research core funding No. P1-0040 and young researcher funding No. PR-10496).

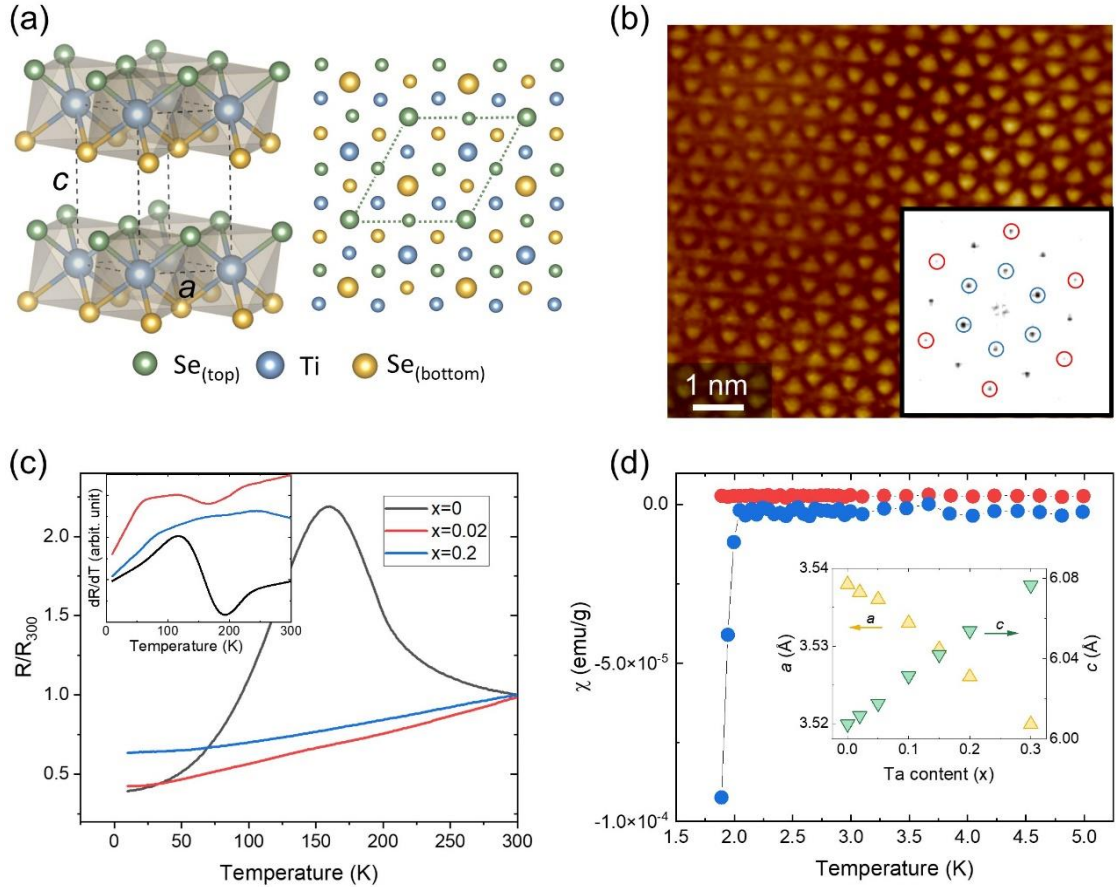

FIG. 1. (a) Atomic structure of 1T-TiSe$_2$. (b) STM image of pristine 1T-TiSe$_2$ with setpoint V$_{tip}$ = 0.1 V, I = 100 pA. The inset shows a Fourier transform of the STM image. Red circles mark lattice peaks and blue circles mark CDW peaks. (c) Resistance measurements of pristine, x = 0.02, and x = 0.2 crystals show a partial (x = 0.02) and full (x = 0.2) suppression of the CDW transition with Ta substitution. The inset shows the derivative of resistance with temperature with x = 0.02 and x = 0.2 curves multiplied by 20 for easier comparison. (d) Magnetic susceptibility at low temperatures shows superconductivity of x=0.2 crystals (blue) with a critical temperature of 2.0 K, while x = 0.02 crystals (red) are not superconducting above 1.9 K. The inset shows lattice parameters of the 1T-Ti$_{1-x}$Ta$_x$Se$_2$ as a function of Ta content obtained from Ref. [26].

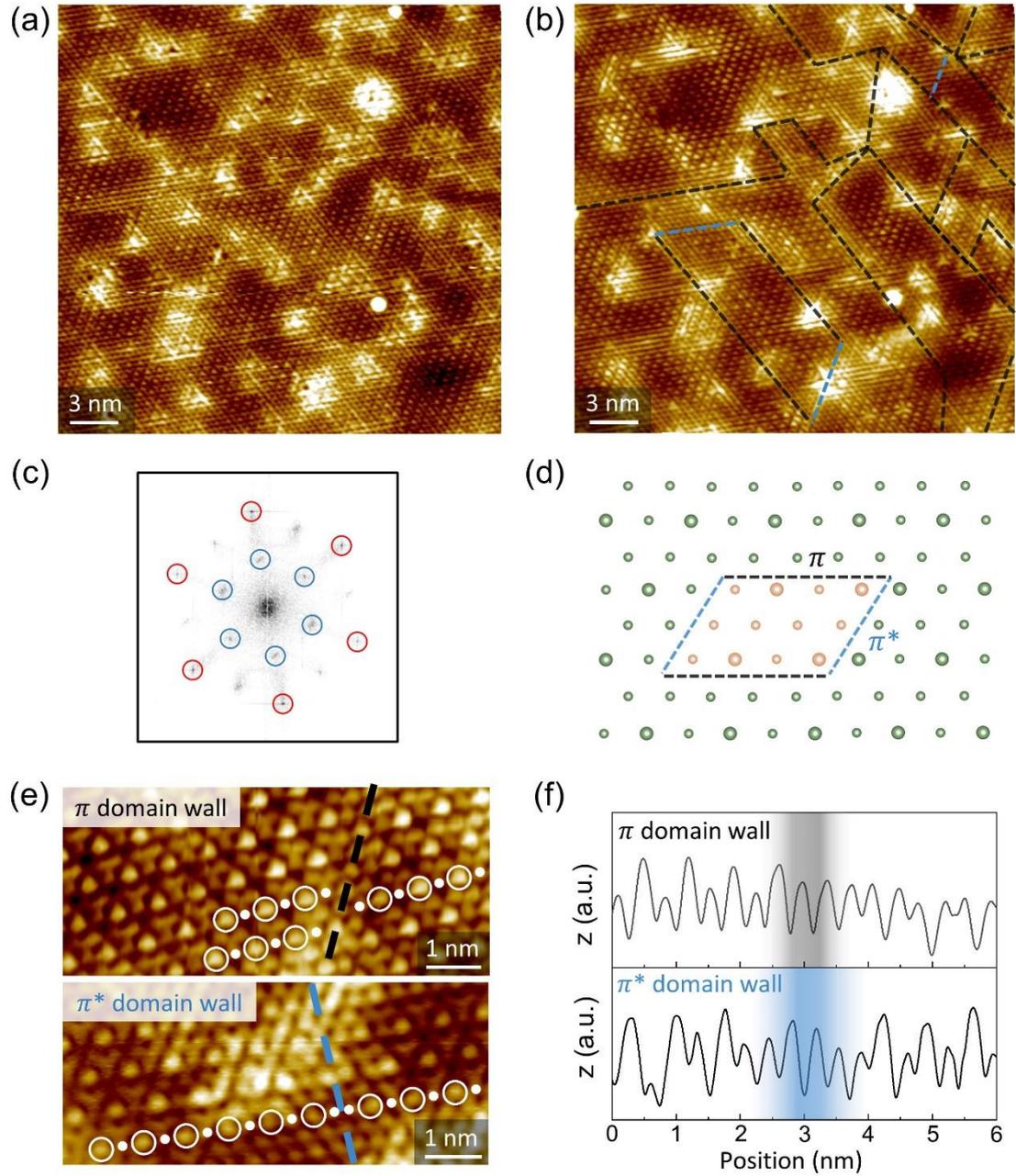

FIG. 2. (a) STM topography image of 1T-Ti$_{0.98}$Ta$_{0.02}$Se$_2$ with setpoint V$_{tip}$ = 0.1 V, I = 100 pA. (b) STM image with V$_{tip}$ = -0.1 V, I = 100 pA with dashed lines marking domain walls. (c) A Fourier transform of the Fig. 2 (c). Red circles mark lattice peaks and blue circles mark CDW peaks. (d) Schematic of 1T-TiSe$_2$ lattice with CDW peak positions emphasized as enlarged atoms. The phase shifted domain inside the larger commensurate state is marked with orange. Both types of π (black dashed line) and π* (blue dashed line) domain walls are shown. (e) V$_{tip}$ = -0.1 V, I = 100 pA and V$_{tip}$ = 0.1 V, I = 100 pA atomically resolved images of both π and π* domain walls. White circles serve as a guide to the eye of CDW ordering and phase slip at the domain wall. (f) Line scans for both π and π* domain walls.

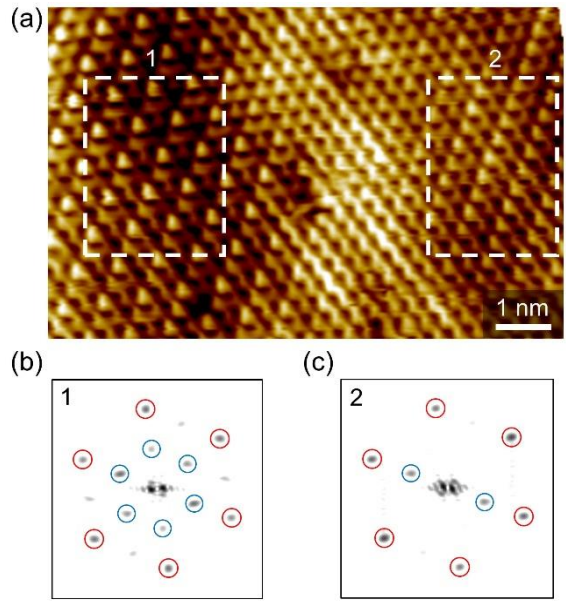

FIG. 3 (a) STM image of 1T-Ti$_{0.98}$Ta$_{0.02}$Se$_2$ with V$_{tip}$ = -0.1 V, I = 70 pA showing both a patch of 3Q CDW (left) and 1Q stripe phase (right). (b) and (c) Fourier transform of patches marked in Fig. 3(a). Red circles mark lattice peaks and blue circles mark CDW peaks.

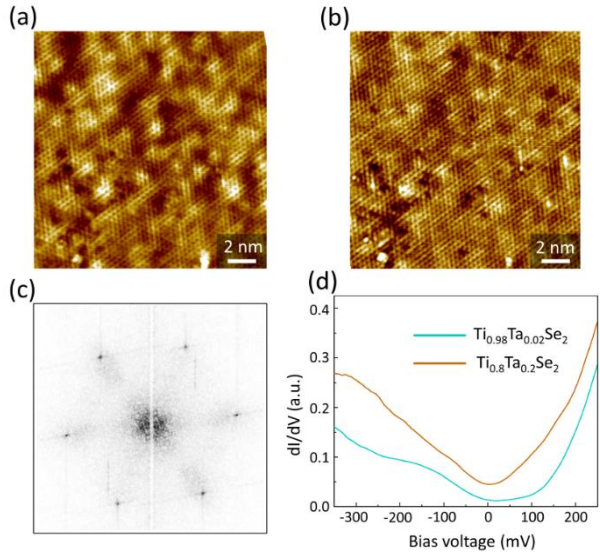

FIG 4. (a) and (b) STM image of 1T-$Ti_{0.8}Ta_{0.2}Se_2$ at $V_{tip}$ = -0.1 V, I = 100 pA and $V_{tip}$ = 0.1 V, I = 100 pA respectively. (c) Fourier transform of STM image in Fig. 4(a). Only lattice peaks are observed and no CDW peaks. (d) Comparison of scanning tunneling spectroscopy measurements of x=0.02 and x=0.2 samples.